\def\vev#1{\langle #1 \rangle}
\def\tr{\mbox{Tr}\,}
\def\ad{\mbox{ad}\,}
\def\ker{\mbox{Ker}\,}

\documentstyle[12pt]{article}
\begin{document}

\title{Quantum Space-time and Classical Gravity}

\author{J. Madore \\
        Laboratoire de Physique Th\'eorique et Hautes 
        Energies\thanks{Laboratoire associ\'e au CNRS, URA D0063}\\
        Universit\'e de Paris-Sud, B\^at. 211, F-91405 Orsay\\       
\and    J. Mourad\\
        GPS, Universit\'e de Cergy-Pontoise\\
        Site de St. Martin, F-95302 Cergy-Pontoise
       }

\date{July, 1996}

\maketitle

\abstract{A method has been recently proposed for defining an arbitrary
number of differential calculi over a given noncommutative associative
algebra. As an example a version of quantized space-time is considered
here. It is found that there is a natural differential calculus using
which the space-time is necessarily flat Minkowski space-time.
Perturbations of this calculus are shown to give rise to non-trivial
gravitational fields.}

\vfill
\noindent
LPTHE Orsay 96/56
\medskip
\eject

\parskip 4pt plus2pt minus2pt

\section{Motivation and Notation}

Since the early days of quantum field theory physicists have been
tempted to introduce some type of lattice structure on space-time to
avoid the appearance of ultraviolet divergences, that is, to fix a
natural value for an ultraviolet cut-off $\Lambda$. One of the
disadvantages of these discrete structures is the fact that they destroy
Lorentz invariance and they can be hardly considered in any respect as
fundamental. It was Snyder (1947) who first had the idea of using
non-commuting coordinates to mimic a discrete structure in a covariant
way. Since then several attempts have been made to continue this initial
effort. We refer to Madore \& Mourad (1996) for a recent review with
historical perspective.  One typically introduces four hermitian 
generators $q^\mu$ of a noncommutative $*$-algebra ${\cal A}$ 
which satisfy commutation relations of the form
$$
[q^\mu, q^\nu] = i \mu_P^{-2} q^{\mu\nu}.                           \eqno(1.1)
$$
The problem lies then with the interpretation of the right-hand side.
One can define a succession of elements $q^{\lambda_1 \dots \lambda_n}$
by the equations
$$
[q^\lambda, q^{\mu\nu}] = i \mu_P^{-1} q^{\lambda\mu\nu}, \qquad
[q^{\lambda}, q^{\mu\nu\rho}] = i \mu_P^{-1} q^{\lambda\mu\nu\rho}  \eqno(1.2)
$$
and so forth. The structure of the algebra ${\cal A}$ is constrained by
the value of these commutators. One possibility, considered by Snyder
(1947), is to choose them so as to form a representation of the Lie
algebra of the de~Sitter group.  A second possibility, considered by
Dubois-Violette \& Madore (Madore 1988, 1995) is to choose them to form
a representation of the conformal algebra.  Recently Doplicher,
Fredenhagen \& Roberts (1994, 1995) have argued that $q^{\mu\nu}$ should
be chosen to lie in the center ${\cal Z}({\cal A})$ of ${\cal A}$, that
the $q$-tensors with more than 2 indices should vanish.  We shall adopt
this as a working hypothesis.  There are 6 independent $q^{\mu\nu}$,
which from (1.1) parameterize symplectic structures on space-time. In
the commutative limit one obtains therefore a space of dimension greater
than four except of course if the $q^{\mu\nu}$ are nilpotent. We shall
not address here the question of the physical significance of the extra
dimensions.

Let $T^{(0)}_{\mu\nu}$ be the bare energy-momentum tensor, including
quantum corrections, of some field theory on space-time, 
Choose some separation of $T^{(0)}_{\mu\nu}$
into a divergent part $T^{(\Lambda)}_{\mu\nu}$ and a regular part
$T^{({\rm Reg})}_{\mu\nu}$ which would remain finite if one were to let
$\Lambda \rightarrow \infty$.  Implicit in what follows is the
assumption that the decomposition can be made so that the singular part
is in some sense universal and independent of the particular (physically
reasonable) field theory one starts with. We write then
$$
T^{(0)}_{\mu\nu} = T^{(\Lambda)}_{\mu\nu} + T^{({\rm Reg})}_{\mu\nu}.
$$
Denote by $\vev{O}_0$ the vacuum-expectation value of an operator $O$. Then
in a quasi-classical approximation, considering the gravitational field
as classical, one can write the Einstein field equations as
$$
G_{\mu\nu} = - \mu_P^{-2} \big(\vev{T^{(\Lambda)}_{\mu\nu}}_0 +
                               \vev{T^{({\rm Reg})}_{\mu\nu}}_0\big).
$$
We shall be here interested in the divergent part of $T^{(0)}_{\mu\nu}$
and we shall neglect the regular term. The field equations become then
$$
G_{\mu\nu} = - \mu_P^{-2} \vev{T^{(\Lambda)}_{\mu\nu}}_0.           \eqno(1.3)
$$
This equation is quite unsatisfactory. One would like to replace it by
an operator equation of the form
$$
G_{\mu\nu} = - \mu_P^{-2} T^{(\Lambda)}_{\mu\nu}                    \eqno(1.4)
$$  
such that
$$
G^{(\infty)}_{\mu\nu} = \lim_{\Lambda \rightarrow \infty} G_{\mu\nu} \eqno(1.5)
$$
is non-vanishing in order to produce a gravitational field which acts as
a regulator but such that
$$
\vev{G^{(\infty)}_{\mu\nu}}_0 = 0                                   \eqno(1.6)
$$
so that the regularizing gravitational field is not classically
observable.  In any case it is reasonable to assume that a
divergence gives rise to a gravitational field and so we have defined
the left arrow of the diagram
$$
\def\normalbaselines{\baselineskip=18pt}
\matrix{
\hbox{Cut-off $\Lambda$} 
&\buildrel \Lambda = \mu_P \over \Longleftrightarrow
&\hbox{Cut-off $\mu_P$}\cr 
\Downarrow && \Uparrow \cr
\hbox{Curvature} &\Longrightarrow &{\cal A}
} 
\def\normalbaselines{\baselineskip=12pt}                            \eqno(1.7)
$$

It is an old idea, due to Pauli and developed by Deser (1957) and others
(Isham {\it et al.} (1971), that perturbative ultraviolet divergences
will one day be regularized by the gravitational field. The possibility
which we would like to explore here is that the mechanism by which this
works is through the introduction of noncommuting `coordinates' such as
the $q^\lambda$. A hand-waving argument can be given (Madore \& Mourad
1995) which allows one to think of the noncommutative structure of
space-time as being due to quantum fluctuations of the light-cone in
ordinary 4-dimensional space-time.  This relies on the existence of
quantum gravitational fluctuations. A purely classical argument based on
the formation of black-holes has been given by Doplicher {\it et al.}
(1995).  In both cases the classical gravitational field $g_{\mu\nu}$ is
to be considered as regularizing the ultraviolet divergences through the
introduction of the quantum structure of space-time.

The right arrow of the Diagram~(1.7) has been discussed, for example, by
Doplicher {\it et al.} The top arrow is a definition.  We wish to
discuss the implications which define the bottom arrow. For this we
consider the following diagram:
$$
\def\normalbaselines{\baselineskip=18pt}
\matrix{
\hbox{Curvature} &\Longleftarrow &\Omega^*({\cal A})\cr 
\Vert && \Downarrow \cr
\hbox{Curvature} &\Longrightarrow &{\cal A}
} 
\def\normalbaselines{\baselineskip=12pt}                            \eqno(1.8)
$$
We shall argue that it can in fact be used to define the bottom arrow,
the same as in (1.7). The right arrow is a mathematical triviality; it
gives a relation between a differential calculus over an algebra and the
algebra itself.  We shall argue that to a certain extent a differential
calculus determines uniquely a curvature in the commutative limit. The
uniqueness will allow us in certain cases to invert the top arrow.  If
we identify the curvature which we so obtain with that which we supposed
was the origin of the structure of the algebra we can claim that
curvature gives rise not only to a noncommutative algebra but also to an
associated differential calculus. As a corollary we have defined
the bottom arrow. Since the differential calculus is not unique we
cannot claim that the curvature depends only on the right-hand side of
(1.1).  That is, although the non-vanishing Planck mass gives rise to
commutation relations, the left-hand side of (1.3) does not depend only
on $\mu_P$.  Equation~(1.3) can therefore not be considered as an
equation for $\Lambda$ in terms of $\mu_P$. Were this the case and were
it possible to use some additional {\it a priori} relation between
$\Lambda$ and $\mu_P$ then (1.3) would become an eigenvalue equation
yielding the mass spectrum in units of $\mu_P$.

In Section~2 we give a general prescription of how one defines
differential calculi based on a set of derivations and we determine the
conditions which the derivations must satisfy for the module of 1-forms
to be a free module and for the module of 2-forms to be non-trivial.  In
Section~3 we associate a metric and a linear connection to each
differential calculus and we argue that the association is unique.  In
Section~4 we give the explicit calculations using a particularly simple
set of derivations and we find that the associated gravitational field
is trivial.  In Sections~5 and 6 we consider differential calculi which are, in
a sense which we shall make precise, small perturbations of this basic
differential calculus and we find that they lead to a rather simple but
non-trivial gravitational field.  In Section~7 we consider a quotient
algebra which could be considered as the noncommutative version of
de~Sitter space. Finally in Section~8 we discuss briefly the definition
of `gauge invariance'.

\section{The differential calculi}

Let ${\cal A}$ be any unital associative $*$-algebra.  Of the many
differential calculi which can be constructed over ${\cal A}$ the
largest is the differential envelope or universal differential calculus
$(\Omega^*_u({\cal A}), d_u)$. Every other differential calculus can be
considered as a quotient of it.  For the definitions we refer, for
example, to the book by Connes (1994). Let $(\Omega^*({\cal A}), d)$ be
another differential calculus over ${\cal A}$. Then there exists a
unique $d_u$-homomorphism $\phi$
$$
\def\normalbaselines{\baselineskip=18pt}
\matrix{
{\cal A} &\buildrel d_u \over \longrightarrow &\Omega_u^1({\cal A})
         &\buildrel d_u \over \longrightarrow &\Omega_u^2({\cal A})
         &\buildrel d_u \over \longrightarrow &\cdots                \cr
\parallel&&\phi_1 \downarrow \phantom{\phi_1}
         &&\phi_2 \downarrow \phantom{\phi_2}                        \cr
{\cal A} &\buildrel d \over \longrightarrow &\Omega^1({\cal A})
         &\buildrel d \over \longrightarrow &\Omega^2({\cal A})
         &\buildrel d \over \longrightarrow &\cdots                  \cr
}
\def\normalbaselines{\baselineskip=12pt}                           \eqno(2.1)
$$
of $\Omega^*_u({\cal A})$ onto $\Omega^*({\cal A})$. It is given by

$$
\phi (d_u f) =  d f.                                               \eqno(2.2)
$$
The restriction $\phi_p$ of $\phi$ to each $\Omega^p_u$ is defined by
$$
\phi_p(f_0 d_u f_1 \cdots d_u f_p) = f_0 df_1 \cdots df_p.
$$

Consider a given algebra ${\cal A}$ and suppose that we know how to
construct an ${\cal A}$-module $\Omega^1({\cal A})$ and an application
$$
{\cal A} \buildrel d \over \longrightarrow \Omega^1({\cal A}).     \eqno(2.3)
$$
Then using (2.1) there is a method of constructing $\Omega^p({\cal A})$
for $p \geq 2$ as well as the extension of the differential.  Since we
know $\Omega_u^1({\cal A})$ and $\Omega^1({\cal A})$ we can suppose that
$\phi_1$ is given. It is the map such that 
$$
\Omega^1({\cal A})=\Omega^1_u({\cal A})/\ker \phi_1.
$$
We must construct $\Omega^2({\cal A})$. The largest consistent choice
would be to set
$$
\Omega^2({\cal A}) =
\Omega_u^2({\cal A})/\overline{d_u \ker \phi_1}               \eqno(2.4)
$$
where 
$$
\overline{d_u \ker \phi_1} = d_u \ker \phi_1
+ \Omega_u^1({\cal A}) \otimes \ker\phi_1
+ \ker\phi_1 \otimes \Omega_u^1({\cal A})
$$
is the bimodule generated by ${d_u \ker \phi_1}$. Since
$$
\Omega^2_u({\cal A})=\Omega^1_u({\cal A})\otimes_{\cal A}
\Omega^1_u({\cal A})
$$
we find that $\Omega^2({\cal A})$ can be written also as
$$
\Omega^2({\cal A}) = 
\Omega^1({\cal A})\otimes_{\cal A}\Omega^1({\cal A}) /{\cal K}     \eqno(2.5)
$$
with
$$
{\cal K} = 
(\phi_1 \otimes \phi_1) (\overline{d_u \ker \phi_1}) =
(\phi_1 \otimes \phi_1) ({d_u \ker \phi_1}).                        \eqno(2.6)
$$
It can happen that 
${\cal K} = \Omega^1({\cal A})\otimes_{\cal A}\Omega^1({\cal A})$, in
which case $\Omega^2({\cal A}) = 0$. 

Let $\pi$ be the projection
$$
\Omega^1({\cal A})\otimes_{\cal A}\Omega^1({\cal A})
\buildrel \pi \over \longrightarrow \Omega^2({\cal A}).             \eqno(2.7)
$$
Then $\pi$ has a right inverse $\iota$, a map
$$
\Omega^2({\cal A}) \buildrel \iota  \over \longrightarrow
\Omega^1({\cal A})\otimes_{\cal A}\Omega^1({\cal A})                \eqno(2.8)
$$
such that $\pi \circ \iota = 1$. This will allow us to identify
$\Omega^2({\cal A})$ as a submodule of 
$\Omega^1({\cal A})\otimes_{\cal A}\Omega^1({\cal A})$. The map 
$\phi_2$ is defined to be the projection of $\Omega_u^2({\cal A})$ 
onto $\Omega^2({\cal A})$ so defined. From the definition of $\pi$
one sees that $\phi_2$ is given by
$$
\phi_2 = \pi \circ (\phi_1 \otimes \phi_1).                         \eqno(2.9)
$$
The wedge product of two elements $\xi$ and $\eta$ in 
$\Omega^1({\cal A})$ is given by $\xi \eta = \pi (\xi \otimes \eta)$.
Let $\xi_u$ be an inverse image of $\xi$ in $\Omega_u^1({\cal A})$. Then
the map $d$ from $\Omega^1({\cal A})$ to $\Omega^2({\cal A})$
can be written in terms of $d_u$ as
$$
d (\phi_1 (\xi_u)) = \phi_2 (d_u \xi_u).                           \eqno(2.10)
$$

Equation~(2.4) defines the largest set of 2-forms consistent with the
constraints on $\Omega^1({\cal A})$.  The procedure can be continued by
iteration to arbitrary order in $p$.  Define for this the map $\psi_p$
$$
\Omega^p_u({\cal A}) \buildrel \psi_p \over \longrightarrow
\bigotimes_1^p \Omega^1({\cal A})                                  \eqno(2.11)
$$
given by $\psi_p = \phi_1 \otimes \cdots \otimes \phi_1$ and define
(Connes 1994)
$$
\Omega^p({\cal A}) =
\Omega^p_u({\cal A})/(\ker \psi_p + 
d_u \ker \psi_{p-1}).                                       \eqno(2.12)
$$
We have then by definition
$$
\ker \phi_p = 
\ker \psi_p + d_u \ker \psi_{p-1}.                   \eqno(2.13)
$$
For example 
$$
\ker \psi_2 = \Omega_u^1({\cal A}) \otimes \ker\phi_1
+ \ker\phi_1 \otimes \Omega_u^1({\cal A})                  \eqno(2.14)
$$
and using (2.13) we find (2.4) as a particular case.  Equation~(2.12)
can be rewritten as
$$
\Omega^p({\cal A}) = 
\bigotimes_1^p \Omega^1({\cal A})/ 
\psi_p (d_u \ker \psi_{p-1})
$$
or in the form
$$
\Omega^p({\cal A}) = 
\bigotimes_1^p \Omega^1({\cal A})/ {\cal K}_p                    \eqno(2.15)
$$
with the ${\cal K}_p$ defined by the recurrence relations
$$
{\cal K}_p = {\cal K}_{p-1} \otimes \Omega^1({\cal A}) + 
\Omega^1({\cal A}) \otimes {\cal K}_{p-1}, \qquad {\cal K}_2 = {\cal K}.
$$
In particular we find the expression
$$
\Omega^3({\cal A}) = 
{\Omega^1({\cal A}) \otimes \Omega^1({\cal A}) \otimes \Omega^1({\cal A})
\over {\cal K} \otimes \Omega^1({\cal A}) +
\Omega^1({\cal A}) \otimes {\cal K}}                             \eqno(2.16)
$$
for the module of 3-forms. 

To initiate the above construction we define the 1-forms using a set of
derivations. We shall suppose that they are interior and exclude
therefore the case where ${\cal A}$ is commutative.  For each integer
$n$ let $\lambda_i$ be a set of $n$ linearly independent
antihermitian elements of ${\cal A}$ and introduce the derivations
$e_i = {\rm ad}\, \lambda_i$.  In general the $e_i$ do
not form a Lie algebra but they do however satisfy commutation relations
as a consequence of the commutation relations of ${\cal A}$. We shall
suppose that if an element of ${\cal A}$ commutes with all of the
$\lambda_i$ then it is in the center ${\cal Z}({\cal A})$ of ${\cal A}$:
$$
e_i f = 0 \Rightarrow f \in {\cal Z}({\cal A}).
$$
Define the map (2.3) by
$$
df (e_i) = e_i \, f = [\lambda_i, f].                              \eqno(2.17)
$$
We shall suppose that there exists a set of $n$ elements $\theta^i$ of
$\Omega^1({\cal A})$ such that
$$
\theta^i (e_j) = \delta^i_j.                                       \eqno(2.18)
$$
In the examples which we consider we shall show that the $\theta^i$
exist by explicit construction.  We shall refer to the set of
$\theta^i$ as a frame or Stehbein.  It commutes with the elements
$f \in {\cal A}$,
$$
f \theta^i = \theta^i f.                                           \eqno(2.19)
$$

The ${\cal A}$-bimodule $\Omega^1({\cal A})$ is generated by all
elements of the form $fdg$ or of the form $(df)g$. Because of the
Leibniz rule these conditions are equivalent. By definition
$$
f dg (e_i) = f e_i \, g, \qquad (dg) f (e_i) =  (e_i \, g) f.
$$
Using the frame we can write these as
$$
f dg = (f e_i g) \theta^i, \qquad (dg) f =  (e_i g) f \theta^i.    \eqno(2.20)
$$
The commutation relations of the algebra constrain the relations
between $f dg$ and $(dg) f$ for all $f$ and $g$.  Because of the
commutation relations of the algebra or, equivalently, because of the
kernel of $\phi_1$ in the quotient (2.4) the $\theta^i$ satisfy in
general commutation relations. With the identification $\iota$ we have
$$
\pi(\theta^i \otimes \theta^j) = 
P^{ij}{}_{kl} \theta^k \otimes \theta^l, \qquad
P^{ij}{}_{kl} \in {\cal Z}({\cal A}).                              \eqno(2.21)
$$
Since $\pi$ is a projection we have
$$
P^{ij}{}_{mn} P^{mn}{}_{kl} =  P^{ij}{}_{kl}                       \eqno(2.22)
$$
and the product $\theta^i \theta^j$ satisfies
$$
\theta^i \theta^j = P^{ij}{}_{kl} \theta^k \theta^l.               \eqno(2.23)
$$
The module ${\cal K}$ is generated by the elements
$(\delta^i_k \delta^j_l - P^{ij}{}_{kl}) \theta^k \otimes \theta^l$.
In one important case which we shall consider the $\theta^i$
anticommute. This corresponds to
$$
P^{ij}{}_{kl} = 
{1\over 2} (\delta^i_k \delta^j_l - \delta^j_k \delta^i_l).        \eqno(2.24)
$$

Define $\theta = - \lambda_i \theta^i$. Then one sees that
$$
df = - [\theta, f]                                                 \eqno(2.25)
$$
and it follows that as a bimodule $\Omega^1({\cal A})$ is generated by
one element.  Under the condition (2.18) the $\Omega^1({\cal A})$ is
free of rank $n$ as a left or right module. It can therefore by
identified with the direct sum of $n$ copies of ${\cal A}$:
$$
\Omega^1({\cal A}) = \bigoplus_1^n {\cal A}.                       \eqno(2.26)
$$
In this representation $\theta^i$ is given by the element of the direct
sum with the unit in the $i^{\rm th}$ position and zero elsewhere.

Any element $\theta_u^i$ of $\Omega_u^1({\cal A})$ can be written in the
form
$$
\theta_u^i = \sum_\alpha f^{(i)}_\alpha \otimes g^{(i)}_\alpha     \eqno(2.27)
$$
with the $f^{(i)}_\alpha$ and $g^{(i)}_\alpha$ elements of ${\cal A}$ such
that
$$
\sum_\alpha f^{(i)}_\alpha g^{(i)}_\alpha = 0.                     \eqno(2.28)
$$
Let $\theta^i$ be the images of $\theta_u^i$ in $\Omega^1({\cal A})$.
Then the condition (2.18) can be rewritten as
$$
\sum_\alpha f^{(i)}_\alpha \lambda_j g^{(i)}_\alpha = \delta^i_j.  \eqno(2.29)
$$
The assumption that $\Omega^1({\cal A})$ is free then is equivalent to
the assumption that these equations have a solution for $f^{(i)}_\alpha$
and $g^{(i)}_\alpha$. 

The $f^{(i)}_\alpha$ and $g^{(i)}_\alpha$ can be used to give an
explicit representation of ${\cal K}$. Introduce the notation
$$
{\cal A} \, [\theta^i_u , {\cal A}] = 
\{ \sum_\alpha f_\alpha [\theta^i_u , g_\alpha] \;\vert\; 
f_\alpha, g_\alpha \in {\cal A} \}.
$$
Then it is easily seen that
$$
\ker \phi_1 = \sum_i {\cal A} \, [\theta^i_u , {\cal A}].   \eqno(2.30)
$$
If we define then the elements $\nu^i$ of 
$\Omega^1({\cal A}) \otimes_{\cal A} \Omega^1({\cal A})$ by
$$
\nu^i = \sum_\alpha f^{(i)}_\alpha \theta \otimes \theta g^{(i)}_\alpha
                                                                   \eqno(2.31)
$$
a short calculation yields the characterization
$$
{\cal K} = \sum_i {\cal A} \, [\nu^i , {\cal A}].                  \eqno(2.32)
$$

If we define $\lambda = (\lambda_1, \dots, \lambda_n)$
then the condition (2.29) can be written in the form
$$
{\cal A} \, [\lambda, {\cal A}] = \bigoplus_1^n {\cal A}.          \eqno(2.33)
$$
More generally let ${\cal B}$ be an ${\cal A}$-bimodule and $\lambda$ an
element of ${\cal B}$. Define a map
$$
{\cal A} \buildrel d \over \longrightarrow {\cal B}                \eqno(2.34)
$$
by $df = [\lambda, f]$. Then we can set
$$
\Omega^1({\cal A}) = {\cal A} \, [\lambda, {\cal A}] \subset {\cal B}.
                                                                   \eqno(2.35)
$$
The choice ${\cal B} = {\cal A} \otimes {\cal A}$ with 
$\lambda = 1 \otimes 1$ yields the universal calculus. We see from this
example that $\lambda$ itself need not be an element of 
$\Omega^1({\cal A})$.

From (2.31) we see that $\pi(\nu^i)$ can be written as
$$
\pi(\nu^i) = 
\sum_\alpha f^{(i)}_\alpha \lambda_j \lambda_k 
P^{jk}{}_{lm} g^{(i)}_\alpha \theta^l \otimes \theta^m.            \eqno(2.36)
$$
But from the identity $[\pi(\nu^i), f] = 0$ we see also that 
$$
\pi(\nu^i) = {1\over 2} F^i{}_{jk} \theta^j \theta^k               \eqno(2.37) 
$$
with $F^i{}_{jk}$ elements in ${\cal Z}({\cal A})$ such that 
$$
P^{jk}{}_{lm} F^i{}_{jk} = F^i{}_{lm}.                             \eqno(2.38)
$$
Using (2.29) it follows that 
$$
\sum_\alpha f^{(i)}_\alpha K_{jk} g^{(i)}_\alpha = 0
$$
if we define $K_{jk}$ by the equation
$$
2 \lambda_l \lambda_m P^{lm}{}_{jk} - 
\lambda_i F^i{}_{jk} - K_{jk} = 0.                                 \eqno(2.39)
$$

The exterior derivative of $\theta^i$ is given by
$$
d\theta^i = \sum_\alpha df^{(i)}_\alpha dg^{(i)}_\alpha
= - [\theta, \theta^i] - \pi(\nu^i).                               \eqno(2.40)
$$
The bracket is a graded bracket.  Multiplying both sides of this
equation by $\lambda_i$ we find the identity
$$
d\theta + \theta^2 = - \theta^2 + \pi(\nu^i \lambda_i).
$$
Using (2.39) we find that this can be written in the form
$$
d\theta + \theta^2 = - {1\over 2} K_{ij} \theta^i \theta^j         \eqno(2.41)
$$
and if we take the exterior derivative of (2.25) we see immediately
that the coefficients $K_{ij}$ must lie in ${\cal Z}({\cal A})$.

The structure elements $C^i{}_{jk}$ are defined by the equation
$$
d\theta^i = 
- {1\over 2} C^i{}_{jk} \theta^j \theta^k.                         \eqno(2.42)
$$
From (2.40) it follows that
$$
C^i{}_{jk} = F^i{}_{jk} - 2 \lambda_l P^{(li)}{}_{jk}.             \eqno(2.43)
$$

We have started from an integer $n$ and a set of $\lambda_i$. The
necessary and sufficient conditions for the existence of the basis
$\theta^i$ are expressed in the Equation~(2.29). If $\Omega^2({\cal A})$
is non-trivial there exists $P^{ij}{}_{kl} \neq 0$ in ${\cal Z}({\cal A})$ 
such that (2.22) and (2.23) are satisfied. Conversely we could have 
started from elements $P^{ij}{}_{kl}$, $F^i{}_{jk}$, $K_{ij}$ in 
${\cal Z}({\cal A})$ and looked for a solution $\lambda_i$ to the
Equation~(2.39). Define
$$
C^{ij}{}_{kl} = \delta^i_k \delta^j_l - 2 P^{ij}{}_{kl}.           \eqno(2.44)
$$
Then from (2.22) we find that 
$$
C^{ij}{}_{kl}C^{kl}{}_{mn} = \delta^i_m \delta^j_n.
$$
From the associativity rule for the product in $\Omega^3({\cal A})$ one
finds that $C^{ij}{}_{kl}$ must satisfy a weak form of the Yang-Baxter
equation.

Quite generally let ${\cal V}$ be an ${\cal A}$-module and $\pi$ a
module morphism
$$
{\cal V} \otimes_{\cal A} {\cal V} 
\buildrel \pi \over \longrightarrow
{\cal V} \otimes_{\cal A} {\cal V}                                 \eqno(2.45) 
$$
with $\pi^2 = \pi$. This is the algebraic generalization of the product
given by (2.7). The equivalent generalization of the product in the 3-forms
is a module morphism
$$
{\cal V} \otimes_{\cal A} {\cal V} \otimes_{\cal A} {\cal V}
\buildrel \pi^\prime \over \longrightarrow
{\cal V} \otimes_{\cal A} {\cal V} \otimes_{\cal A} {\cal V}       \eqno(2.46) 
$$
with $\pi^{\prime 2} = \pi^\prime$. One has then
$$
\ker \pi^\prime = \ker \pi \otimes {\cal V} +
{\cal V} \otimes \ker \pi.                                  \eqno(2.47)
$$
For the product to be non-trivial we must require that 
$\pi^\prime \neq 0$. Since we have
$$
\pi^\prime (\ker \pi \otimes {\cal V}) = 0, \qquad
\pi^\prime ({\cal V} \otimes \ker \pi) = 0,
$$
there must exist two morphisms $\phi^\prime$, $\phi^{\prime\prime}$ of
${\cal V} \otimes_{\cal A} {\cal V} \otimes_{\cal A} {\cal V}$ into
itself such that
$$
\pi^\prime = \phi^\prime \circ \pi_{12}, \qquad 
\pi^\prime = \phi^{\prime\prime} \circ \pi_{23}.
$$
We have here used the standard convention of setting 
$\pi_{12} = \pi \otimes 1$ and $\pi_{23} = 1 \otimes \pi$.
The associativity rule becomes the compatibility condition
$$
\phi^\prime \circ \pi_{12} = \phi^{\prime\prime} \circ \pi_{23}.   \eqno(2.48)
$$
In particular if $C^{ij}{}_{kl}$ satisfies the Yang-Baxter condition
$$
C_{23} \circ C_{12} \circ C_{23} = C_{12} \circ C_{23} \circ C_{12}
$$
then one can set 
$$
\phi^\prime = - {1 \over 3}(1 - 4 \pi_{12} \circ \pi_{23}), \qquad 
\phi^{\prime\prime} = - {1 \over 3}(1 - 4\pi_{23} \circ \pi_{12}). \eqno(2.49)
$$
However more general solutions to (2.48) do exist.
 
The $F^i{}_{jk}$ must satisfy a set of modified Jacobi identities.  If
we choose the $\lambda_i$ so that the $e_i$ are a basis of the Lie
algebra of all derivations of a matrix algebra (Dubois-Violette {\it et
al.} 1989) then one can choose $P^{ij}{}_{kl} = \delta^i_k\delta^j_l$. 
If a smaller Lie algebra (Madore 1995) is chosen then $P^{ij}{}_{kl}$ is
given by (2.24), $K_{ij} = 0$ and the $F^i{}_{jk}$ are equal to the
structure constants of the Lie algebra. An example with $K_{ij} \neq 0$
is to be found in Dubois-Violette {\it et al.} (1996b). Examples with
$F^i{}_{jk} = 0$ and $K_{ij} = 0$ are given in Dimakis \& Madore (1996).
If $P^{ij}{}_{kl}$ is given by (2.24) with a plus instead of a minus
sign, $n$ is even and $F^i{}_{jk} = 0$ then a solution to (2.39) is
given by Dirac matrices.  If also $K_{ij} = 0$ then a solution is given
by `super-coordinates'. In these two cases the 1-forms $\theta^i$ commute.

\section{The linear connections}

The definition of a connection as a covariant derivative was given an
algebraic form in the Tata lectures by Koszul (1960) and generalized to
noncommutative geometry by Karoubi (1981) and Connes (1986, 1994). We
shall use here the expressions `connection' and `covariant derivative'
synonymously.  A `bimodule connection' is a connection on a general
bimodule ${\cal M}$, which satisfies a left and right Leibniz rule. In
the particular case where ${\cal M}$ is the module of 1-forms we shall
speak of a `linear connection'.

Let ${\cal A}$ be an arbitrary algebra and $(\Omega^*({\cal A}) ,d)$ a
differential calculus over ${\cal A}$. One defines a left 
${\cal A}$-connection on a left ${\cal A}$-module ${\cal H}$ as a
covariant derivative
$$
{\cal H} \buildrel D \over \rightarrow 
\Omega^1({\cal A}) \otimes_{\cal A} {\cal H}                      \eqno(3.1) 
$$ 
which satisfies the left Leibniz rule
$$
D (f \psi) =  df \otimes \psi + f D\psi                           \eqno(3.2)
$$
for arbitrary $f \in {\cal A}$. This map has a natural extension
$$
\Omega^*({\cal A}) \otimes_{\cal A} {\cal H} 
\buildrel D \over \longrightarrow
\Omega^*({\cal A}) \otimes_{\cal A} {\cal H}                      \eqno(3.3)
$$
given, for $\psi \in {\cal H}$ and $\alpha \in \Omega^n({\cal A})$,
by
$$
D (\alpha \psi) = d\alpha \otimes \psi +  (-1)^n \alpha D \psi.
$$
The operator $D^2$ is necessarily left-linear.

A covariant derivative on the module $\Omega^1({\cal A})$ must satisfy
(3.2). But $\Omega^1({\cal A})$ has also a natural structure as a right
${\cal A}$-module and one must be able to write a corresponding right
Leibniz rule in order to construct a bilinear curvature. Quite generally
let ${\cal M}$ be an arbitrary bimodule. Consider a covariant derivative
$$
{\cal M} \buildrel D \over \rightarrow 
\Omega^1({\cal A}) \otimes_{\cal A} {\cal M}                       \eqno(3.4) 
$$ 
which satisfies both a left and a right Leibniz rule.  In order to
define a right Leibniz rule which is consistent with the left one, it
was proposed by Mourad (1995), by Dubois-Violette \& Michor (1996) and
by Dubois-Violette \& Masson (1996) to introduce a generalized
permutation
$$
{\cal M} \otimes_{\cal A} \Omega^1({\cal A})
\buildrel \sigma \over \longrightarrow
\Omega^1({\cal A}) \otimes_{\cal A} {\cal M}.
$$
The right Leibniz rule is given then as
$$
D(\xi f) = \sigma (\xi \otimes df) + (D\xi) f                     \eqno(3.5)
$$
for arbitrary $f \in {\cal A}$ and $\xi \in {\cal M}$. The purpose of
the map $\sigma$ is to bring the differential to the left while
respecting the order of the factors. It is necessarily bilinear. 
Consider the case 
$$
{\cal M} = \Omega^1({\cal A})
$$
and let $\pi$ be the projector defined by (2.7).  It was shown by Mourad
(1995) and by Dubois-Violette {\it et al.} (1995) that a necessary as
well as sufficient condition for torsion to be right-linear is that
$\sigma$ satisfy the consistency condition
$$
\pi \circ (\sigma + 1) = 0.                                        \eqno(3.6) 
$$
We define a bimodule ${\cal A}$-connection to be the couple 
$(D, \sigma)$. 
Using the fact that $\pi$ is a projection one sees that the most general
solution to the constraint (3.6) is given by
$$
1 + \sigma = ( 1 - \pi) \circ \tau                                 \eqno(3.7)
$$
where $\tau$ is an arbitrary map
$$
\Omega^1({\cal A}) \otimes \Omega^1({\cal A})
\buildrel \tau \over \longrightarrow
\Omega^1({\cal A}) \otimes \Omega^1({\cal A}).                     \eqno(3.8)
$$
If we choose $\tau = 2$ then we find $\sigma = 1 - 2 \pi$ and 
$\sigma^2 = 1$. The eigenvalues of $\sigma$ are then equal to $\pm 1$.  

There is at the moment no general consensus of the correct definition of
the curvature of a bimodule connection. The problem is that the operator
$D^2$ need not in general be right-linear.  Nevertheless in the
particular cases of interest to us here, with a module of 1-forms which
is free and has a special basis such that (2.19) is satisfied, the
ordinary definition of curvature is quite satisfactory. We refer to
Dubois-Violette {\it et al.} (1996b) or to Dimakis (1996) for 
recent discussions.

This general formalism can be applied in particular to the differential
calculi which we have constructed in the previous section.  Since
$\Omega^1({\cal A})$ is a free module the maps $\sigma$ and $\tau$ 
can be defined by their action on the basis elements:
$$
\sigma (\theta^i \otimes \theta^j) = 
S^{ij}{}_{kl} \theta^k \otimes \theta^l,   \qquad
\tau (\theta^i \otimes \theta^j) = 
T^{ij}{}_{kl} \theta^k \otimes \theta^l.                           \eqno(3.9)
$$
By the sequence of identities
$$
f S^{ij}{}_{kl} \theta^k \otimes \theta^l = 
\sigma (f \theta^i \otimes \theta^j) = 
\sigma (\theta^i \otimes \theta^j f) =
S^{ij}{}_{kl} f \theta^k \otimes \theta^l                          \eqno(3.10)
$$
and the corresponding ones for $T^{ij}{}_{kl}$ we conclude that the
coefficients $S^{ij}{}_{kl}$ and $T^{ij}{}_{kl}$ must lie in 
${\cal Z}({\cal A})$. From (3.7) the most general form for 
$S^{ij}{}_{kl}$ is
$$
S^{ij}{}_{kl} = 
(\delta^i_m \delta^j_n - P^{ij}{}_{mn}) T^{mn}{}_{kl} 
- \delta^i_k \delta^j_l.                                           \eqno(3.11)
$$

Since $\Omega^1({\cal A})$ is a free module a covariant
derivative can be defined by its action on the basis elements:
$$
D\theta^i = - \omega^i{}_{jk} \theta^j \otimes \theta^k.           \eqno(3.12)
$$
The coefficients here are elements of the algebra.  The torsion 2-form
is defined as usual as
$$
\Theta^i = d \theta^i - \pi \circ D \theta^i.
$$
There is a natural covariant derivative $D_{(0)}$ (Dubois-Violette {\it
et al.} 1996b) given by
$$
D_{(0)} \theta^i = - \theta \otimes \theta^i 
+ \sigma (\theta^i \otimes \theta) - \nu^i,                        \eqno(3.13)
$$
which is torsion-free by (2.40). The corresponding coefficients are given
by
$$
\omega_{(0)}{}^i{}_{jk} = 
\lambda_l (S^{il}{}_{jk} - \delta^l_j \delta^i_k) +
{1 \over 2} F^i{}_{jk}.                                            \eqno(3.14)
$$
The most general $D$ is of the form
$$
D = D_{(0)} + \chi                                                 \eqno(3.15)
$$
where $\chi$ is an arbitrary bimodule morphism
$$
\Omega^1({\cal A}) \buildrel \chi \over \longrightarrow
\Omega^1({\cal A}) \otimes \Omega^1({\cal A}).                     \eqno(3.16)
$$
If we write
$$
\chi (\theta^i) = - \chi^i{}_{jk} \theta^j \otimes \theta^k        \eqno(3.17)
$$
then by an argument similar to (3.10) we conclude that
$$
\chi^i{}_{jk} \in {\cal Z}({\cal A}).                              \eqno(3.18)
$$
In general a covariant derivative is torsion-free provided the condition
$$
\omega^i{}_{jk} - \omega^i{}_{lm} S^{lm}{}_{jk} = C^i{}_{jk}       \eqno(3.19)
$$
is satisfied. The covariant derivative (3.15) is torsion free if and only
if
$$
\pi \circ \chi = 0.                                                \eqno(3.20) 
$$

One can define a metric by the condition
$$
g(\theta^i \otimes \theta^j) = g^{ij}                              \eqno(3.21)
$$
where the coefficients $g^{ij}$ are elements of the algebra.
To be well defined on all elements of the tensor product 
$\Omega^1({\cal A}) \otimes_{\cal A} \Omega^1({\cal A})$ the metric must
be bilinear and by the sequence of identities
$$
f g^{ij} = g(f \theta^i \otimes \theta^j) 
= g(\theta^i \otimes \theta^j f) = g^{ij} f                        \eqno(3.22)
$$
one concludes that the coefficients must lie in ${\cal Z}({\cal A})$.
This restriction plays an important role in the unicity argument which
allows us to invert the top arrow of Diagram~(1.8). In the commutative
limit the $g^{ij}$ cannot be functions of the coordinates. In ordinary
geometry an equivalence class of moving frames determines a metric and
all equivalence classes correspond to the same differential calculus,
the ordinary de~Rham differential calculus. The choice of differential
calculus does not fix the metric. In the noncommutative case on the other
hand, as we have defined it, each differential calculus determines a
Stehbein and thereby a metric. In the commutative limit all of the
noncommutative differential calculi are either singular, if $n$ is not
equal to the classical dimension of the manifold, or have a common
limit. The moving frame however and the associated metric remain however
as a shadow of the noncommutative structure.

The covariant derivative (3.12) is compatible with the metric
(Dubois-Violette {\it et al.} 1995) if and only if
$$
\omega^i{}_{jk} + \omega_{kl}{}^m S^{il}{}_{jm} = 0.              \eqno(3.23)
$$
When $F^i{}_{jk} = 0$ the condition that (3.14) be metric compatible can
be written as
$$
S^{im}{}_{ln} g^{np} S^{kj}{}_{mp} = g^{ik} \delta^j_l.           \eqno(3.24)
$$
The metric we have chosen is not symmetric with respect to $\sigma$.
That is
$$
g^{ij} \neq S^{ij}{}_{kl} g^{kl}
$$
in general. If one wishes to find a metric symmetric in the above sense
then one must consider (3.24) as an equation for $S$ and the metric and
add the additional equation
$$
g^{ij} = S^{ij}{}_{kl} g^{kl}.
                                                                   \eqno(3.25)
$$
The system (3.24), (3.25), if it has a solution, would yield a symmetric
metric with a compatible connection.

Since we are primarily interested in the first-order effects in the
commutative limit we can identify the curvature with the operator $D^2$.
We set as usual then
$$
D^2 \theta^i = - {1 \over 2} R^i{}_{jkl}
\theta^k \theta^l \otimes \theta^j.                                \eqno(3.26)
$$
Since $D^2$ is not necessarily right-linear as an operator the last of
the equivalent of the sequence of identities (3.10) is not valid and we
cannot conclude that the coefficients $R^i{}_{jkl}$ necessarily lie in
the center of the algebra.

\section{The basic calculus}

In this section we shall return to the `space-time' algebra defined in
Section~1 and we shall suppose, with Doplicher {\it et al.}, that the
$q^{\mu\nu}$ lie in the center of the algebra.  This permits us to
suppose further that the matrix $q^{\mu\nu}$ has an inverse
$q^{-1}_{\lambda\mu}$:
$$
q^{-1}_{\lambda\mu}q^{\mu\nu} = \delta^\nu_\lambda.
$$
We shall use this inverse to lower the indices of the generators $q^\mu$:
$$
\tilde q_\lambda = \mu_P^2 q^{-1}_{\lambda\mu} q^\mu.
$$
A natural choice of $n$ is $n=4$ and a natural choice of $\lambda_\mu$ 
is given by
$$
\lambda_\mu = - i \tilde q_\mu.                                     \eqno(4.1)
$$
The associated derivations defined in Section~2 satisfy then
$$
e_\mu q^\lambda = \delta^\lambda_\mu                                \eqno(4.2)
$$
and it follows that
$$
[e_\mu, e_\nu] = 0.                                                 \eqno(4.3)
$$

From (4.2) it follows that
$$
\theta^\lambda = dq^\lambda, \qquad 
\theta = i \tilde q_\lambda dq^\lambda                              \eqno(4.4) 
$$
from which we deduce that
$$
P^{\mu\nu}{}_{\rho\sigma} = 
{1\over 2} (\delta^\mu_\rho \delta^\nu_\sigma - 
\delta^\nu_\rho \delta^\mu_\sigma), \qquad 
F^\lambda{}_{\mu\nu} = 0, \qquad
K_{\mu\nu} = i \mu_P^2 q^{-1}_{\mu\nu}.                             \eqno(4.5)
$$
One can interpret $\theta$ as a connection on a trivial bundle with the
unitary elements of the algebra as structural group (Dubois-Violette
{\it et al.} 1989, 1990).  We see from the above formula for
$K_{\mu\nu}$ that $q^{\mu\nu}$ is related to the corresponding
curvature. This is the noncommutative analogue of the classical result
of mechanics which interprets the symplectic 2-form as the curvature of
a line bundle.

From the commutation relations (2.19) one finds that the $\theta^\lambda$
anticommute. A possible form for $\sigma$ is given therefore by
$$
S^{\mu\nu}{}_{\rho\sigma} = \delta^\nu_\rho \delta^\mu_\sigma.
                                                                    \eqno(4.6)
$$
From (3.14) we see that in this case the coefficients of the connection
necessarily lie in the center of the algebra. From (3.6) we see that 
the most general $S^{\mu\nu}{}_{\rho\sigma}$ must satisfy the 
constraint
$$
S^{\mu\nu}{}_{[\rho\sigma]} +
\delta^\mu_{[\rho} \delta^\nu_{\sigma]} = 0.                        \eqno(4.7)
$$

The most general $\sigma$ is defined by a solution to the
Equations~(3.24) and (4.7).  If we restrict the $g_{\lambda\mu}$ to be
the components of the Minkowski metric then the unique solution is given
by (4.6). The Minkowski metric is then symmetric also with respect to
$\sigma$. From (3.19) and (3.23) we see that if we require that the
torsion vanish then we have
$$
\omega^\lambda{}_{\mu\nu} = 0.                                      \eqno(4.8)
$$
The space-time is therefore a noncommutative version of Minkowski space
and the right-hand side of Equation~(1.3) must vanish.

It is of interest to notice that Equation~(4.2) defines a derivation of
the algebra whatever the form of the matrix $q^{\mu\nu}$. The derivation
is inner if the matrix is invertible; otherwise it is outer. Let 
$\theta^\lambda$ be a set of Grassmann variables. Define
$$
q^\lambda = x^\lambda + \mu^{-1}_P \theta^\lambda, \qquad
q^{\mu\nu} = - 2 i \theta^\mu \theta^\nu.                         \eqno(4.9)
$$
Then (1.1) is satisfied and $q^{\mu\nu} \in {\cal Z}({\cal A})$. In this
case the matrix $q^{\mu\nu}$ is not invertible; it is in fact nilpotent. 

Notice also that in this case the center ${\cal Z}({\cal A})$ is
nontrivial and in fact it is possible to impose that it be a smooth
6-dimensional manifold with the $q^{\mu\nu}$ as coordinates. We can, with
Doplicher {\it et al.}, impose the conditions
$$
q_{\mu\nu} q^{\mu\nu} = 0, \qquad                                  \eqno(4.10)
$$
as well as
$$
\epsilon_{\mu\nu\rho\sigma} q^{\mu\nu} q^{\rho\sigma} = 12,
$$
which is equivalent to
$$
q^{-1}_{\mu\nu} = 
- {1\over 3} \epsilon_{\mu\nu\rho\sigma}q^{\rho\sigma}.            \eqno(4.11)
$$
The normalization is chosen for later convenience.  The manifold can
then be reduced to 4 dimensions. In the limit $\mu_P \rightarrow 0$ we
have in fact a structure which can be regarded as a 4-dimensional
manifold with a non-commutative extension \`a la Kaluza-Klein similar to
the structures which are mentioned, for example, in Madore \& Mourad
(1996). In the limit $\mu_P \rightarrow \infty$ the structure can be
considered to be that of an ordinary space-time with an extra 4- or
6-dimensional factor in which the Poisson structure defined by the
commutator (1.1) takes its values (Dubois-Violette {\it et al.} (1996a).
An element of ${\cal Z}({\cal A})$ can in no way correspond to a
function on space-time in the commutative limit.  Also if $f$ is an
element of ${\cal A}$ such that $e_\mu f = 0$ then we can only conclude
that $f$ is an arbitrary function of the $q^{\mu\nu}$; we cannot
conclude that it is proportional to the identity.  We regard the
non-trivial center as something which is to be eventually eliminated for
example by choosing $q^{\mu\nu}$ not to lie in the center.  To simplify
the calculations we shall suppose that the matrix of coefficients
$g^{\mu\nu}$ of the metric is symmetric in the ordinary sense of the
word we shall impose the condition that it be equal to the matrix of
components of the ordinary Minkowski metric.

\section{Variations of the calculus}

We shall now check the stability of the result (4.8) under the
perturbation of the differential calculus. Introduce four arbitrary
`small' elements $f^\lambda$ of ${\cal A}$ and define
$$
\tilde f_\lambda = \mu_P^2 q^{-1}_{\lambda\mu} f^\mu.              \eqno(5.1)
$$
Then the elements 
$$
\lambda^\prime_\mu = - i (\tilde q_\mu +  \tilde f_\mu)            \eqno(5.2)
$$
are `near' to (4.1). In general, unless the condition (2.39) is
satisfied, $\Omega^2({\cal A}) = 0$ and the curvature will vanish.
Impose the condition (2.39) and let $P_{(1)}^{\rho\sigma}{}_{\mu\nu}$,
$F_{(1)}^\lambda{}_{\mu\nu}$ and $K_{(1)}{}_{\mu\nu}$ be the first-order
perturbations respectively of the coefficients.  By simple dimensional
arguments one can argue that $P_{(1)}^{\mu\nu}{}_{\rho\sigma}$ must
vanish. In fact it must tend to zero when the Planck mass tends to
infinity but on the other hand it is without dimension and therefore
cannot depend on the Planck mass.  Therefore we set
$$
P_{(1)}^{\mu\nu}{}_{\rho\sigma} = 0.                               \eqno(5.3)
$$
Using (4.5) we find that the linearization of (2.39) yields the equation
$$
[\tilde q_\mu, \tilde f_\nu] - [\tilde q_\nu, \tilde f_\mu] = 
i F_{(1)}^\lambda{}_{\mu\nu} \tilde q_\lambda - K_{(1)}{}_{\mu\nu}.\eqno(5.4) 
$$
Let $k_\mu$ be an arbitrary `small' 4-vector with the dimension of mass.
Then a solution is given by
$$
\tilde f_\lambda = k_\mu q^\mu \tilde q_\lambda                    \eqno(5.5)
$$
and
$$
F_{(1)}^\lambda{}_{\mu\nu} = 
k^{\phantom{\lambda}}_{[\mu} \delta^\lambda_{\nu]} +
2 q^{-1}_{\mu\nu} q^{\lambda\sigma} k_\sigma, \qquad
K_{(1)}{}_{\mu\nu} = 0.                                            \eqno(5.6) 
$$
The corresponding frame is given by
$$
\theta^{\prime\lambda} = (1 - k_\rho q^\rho) dq^\lambda + 
q^{\lambda\rho} k_\rho q^{-1}_{\mu\sigma} q^\sigma dq^\mu.         \eqno(5.7)
$$
It will generate a new differential calculus 
$\Omega^{\prime *}({\cal A})$ which will be in general different 
from $\Omega^*({\cal A})$. Using (2.42) or (2.43) we find that 
$$
C_{(1)}^\lambda{}_{\mu\nu} = F_{(1)}^\lambda{}_{\mu\nu}.           \eqno(5.8)
$$

The frame (5.7) explicitly breaks Lorentz invariance through the vector
$k_\lambda$.  This is quite natural since there is now present a
gravitational field in the commutative limit. What is less desirable is
the dependence on the extra `coordinates' $q^{\mu\nu}$. This can be
eliminated by introducing a Lorentz-invariant probability measure on the
space of $q^{\mu\nu}$ with
$$
\vev{q^{\mu\nu} q^{\rho\sigma}} = \epsilon^{\mu\nu\rho\sigma}.     \eqno(5.9)
$$
The normalization in (4.11) was chosen so that this equation is
compatible with the condition $\vev{1} = 1$. We have then
$$
\vev{\theta^{\prime\lambda}} = dq^\lambda - q^\lambda k_\mu dq^\mu \eqno(5.10)
$$
and
$$
\vev{F_{(1)}^\lambda{}_{\mu\nu}} = 
- k^{\phantom{\lambda}}_{[\mu} \delta^\lambda_{\nu]}.             \eqno(5.11)
$$
The particular, simple form of (5.10) and (5.11) is due to the choice of
normalization of the measure on the space of $q^{\mu\nu}$.

Let $D^\prime$ be a covariant derivative and define the coefficients
$\omega^{\prime\lambda}{}_{\mu\nu}$ by the equation
$$
D^\prime\theta^{\prime\lambda} = - \omega^{\prime\lambda}{}_{\mu\nu} 
\theta^{\prime\mu} \otimes \theta^{\prime\nu}
$$
equivalent to (3.12). Because of (4.8) we have 
$\omega^{\prime\lambda}{}_{\mu\nu} = \omega_{(1)}^\lambda{}_{\mu\nu}$ and
we can write to lowest order
$$
D^\prime\theta^{\prime\lambda} = 
- \omega^\lambda_{(1)\mu\nu} 
\theta^{\prime\mu} \otimes \theta^{\prime\nu}.                    \eqno(5.12)
$$
To extend the covariant derivative to the entire module of 1-forms we
shall need the expression for the perturbed value $\sigma^\prime$ of 
$\sigma$. We define the coefficients $S^{\prime\mu\nu}{}_{\rho\sigma}$ 
by the equation
$$
\sigma^\prime (\theta^{\prime\mu} \otimes \theta^{\prime\nu}) =
S^{\prime\mu\nu}{}_{\rho\sigma} 
\theta^{\prime\rho} \otimes \theta^{\prime\sigma}
$$
equivalent to (3.12). We expand
$$
S^{\prime\mu\nu}{}_{\rho\sigma} = \delta^\nu_\rho \delta^\mu_\sigma
+ S_{(1)}^{\mu\nu}{}_{\rho\sigma}.                                 \eqno(5.13)
$$
Each choice of $S_{(1)}^{\mu\nu}{}_{\rho\sigma}$ corresponds to a
definite choice of covariant derivative. As above, by simple dimensional
arguments one can argue that
$$
S_{(1)}^{\mu\nu}{}_{\rho\sigma} = 0.                               \eqno(5.14)
$$
Therefore from (3.14) and a proper choice of $\chi$ so that (3.23) is 
satisfied we find that the torsion-free metric connection is given by
$$
\omega_{(1)}^\lambda{}_{\mu\nu} = 
{1\over 2} (F_{(1)}^\lambda{}_{\mu\nu} 
- F_{(1)}{}_{\nu}{}^\lambda{}_\mu 
+ F_{(1)}{}_{\mu\nu}{}^\lambda).                                   \eqno(5.15)
$$
The curvature of the covariant derivative defined by this expression is
essentially constant as far as `space-time' is concerned. From this
point of view it is not particularly interesting. Using (5.11) we find
that
$$
\vev{\omega_{(1)}^\lambda{}_{\mu\nu}} = 
{1\over 2} (k^\lambda g_{\mu\nu} - k_\mu \delta^\lambda_\nu).      \eqno(5.16)
$$

To lowest order we find the expression
$$
R^\mu{}_{\nu\rho\sigma} = 
\omega_{(1)}^\mu{}_{\rho\tau} \omega_{(1)}^\tau{}_{\sigma\nu} -
\omega_{(1)}^\mu{}_{\sigma\tau} \omega_{(1)}^\tau{}_{\rho\nu} -
\omega_{(1)}^\mu{}_{\tau\nu} C_{(1)}^\mu{}_{\rho\sigma}           \eqno(5.17)
$$
for the components of the curvature. It is not of particular interest to
give an expression for the expectation value
$\vev{R^\mu{}_{\nu\rho\sigma}}$ since it will depend critically on the
probability measure. However the expectation value of the 
Einstein tensor must be of the form 
$$
\vev{G_{\mu\nu}} = a k_\mu k_\nu + b k^2 g_{\mu\nu}                \eqno(5.18)
$$
where $a$ and $b$ are dimensionless constants which depend on the
details of the probability measure. According to the logic of the
diagram (1.8) this expression is to be substituted for the left-hand
side of Equation~(1.3). At the present preliminary stage of the
understanding of the relation between the differential calculus and the
resulting curvature it is premature to consider this equation further.

\section{Variations of the algebra}

Another way to obtain a non-vanishing gravitational field is to vary the
structure of the algebra ${\cal A}$. We introduce 6 `small' elements
$q_{(1)}^{\mu\nu}$ of ${\cal A}$ and define
$$
q^{\prime\mu\nu} = q^{\mu\nu} + q_{(1)}^{\mu\nu}.                 \eqno(6.1)
$$
We have then
$$
q^{\prime -1}_{\mu\nu} = q^{-1}_{\mu\nu} + 
q^{-1}_{\mu\rho} q^{-1}_{\nu\sigma} q_{(1)}^{\rho\sigma}.         \eqno(6.2)
$$
Since we are here primarily interested in the effect of varying the
structure of the algebra we keep the basic calculus and set
$$
\lambda^\prime_\mu = \lambda_\mu.                                  \eqno(6.3)
$$
From (1.2) we find
$$
[q^\lambda, q^{\prime\mu\nu}] = [q^\lambda, q_{(1)}^{\mu\nu}] = 
i \mu_P^{-1} q_{(1)}^{\lambda\mu\nu}.
$$
The simplest generalization of the basic algebra is obtained by
supposing that $q_{(1)}^{\lambda\mu\nu}$ lies in the center of ${\cal A}$. 
This is the extended model of Doplicher {\it et al}. If we impose
this condition we can choose
$$
q_{(1)}^{\mu\nu} = - \mu_P^{-1} q_{(1)}^{\lambda\mu\nu} \tilde q_\lambda.
                                                                  \eqno(6.4)
$$
We have then
$$
e^\prime_\mu q^\nu = e_\mu q^\nu = \delta_\mu^\nu + 
\mu_P^{-1} q^{-1}_{\mu\rho} q_{(1)}^{\nu\rho\sigma} \tilde q_\sigma.
                                                                   \eqno(6.5)
$$
Using (4.5) we find that the linearization of (2.39) yields now the
solution
$$
F_{(1)}^\lambda{}_{\rho\sigma} = 
\mu_P q^{-1}_{\rho\mu} q^{-1}_{\sigma\nu}
\big(q_{(1)}^{\lambda\mu\nu} - q_{(1)}^{\mu\nu\lambda} + 
q_{(1)}^{\nu\mu\lambda}\big).                                       \eqno(6.6)
$$
The corresponding frame is given by
$$
\theta^{\prime\lambda} = dq^\lambda - \mu_P^{-1} q^{-1}_{\mu\rho} 
q_{(1)}^{\lambda\rho\sigma}\tilde q_\sigma dq^\mu.                  \eqno(6.7)
$$

\section{de Sitter space}

To define a noncommutative version of a space which is not topologically
trivial we shall use elementary techniques from classical geometry as
well as from the quantum mechanics of constrained hamiltonian systems. A
non-trivial topological manifold $V$ can be defined by its imbedding in
a flat space of sufficiently high dimension. The algebra 
${\cal B} = {\cal C}(V)$ of continuous functions on $V$ can be
identified with the algebra ${\cal A}$ of all continuous functions on
the enveloping space modulo the ideal ${\cal I}$ of continuous functions
which vanish on $V$: ${\cal B} = {\cal A}/{\cal I}$. As a first attempt
to define a noncommutative version of de~Sitter space we set ${\cal A}$
equal to a $*$-algebra generated by a set of elements $q^i$, 
$0 \leq i \leq 4$, which satisfy commutation relations similar to those
given by (1.1):
$$
[q^i, q^j] = i \mu_P^{-2} q^{ij}.                                   \eqno(7.1)
$$
We introduce a Minkowski-signature metric with components $g_{ij}$ and
we define ${\cal I}$ to be the 2-sided ideal generated by the
element
$$
c_1 = g_{ij} q^i q^j - r^2                                          \eqno(7.2)
$$
where $r^2$ is a real constant.  Two problems present themselves
immediately. The $q^{ij}$ cannot be invertible; it would otherwise
define a symplectic form on an odd-dimensional manifold. Using the
commutation relations (7.1) one sees also that the 2-sided ideal
generated by the element $c_1$ is the entire algebra and therefore
${\cal B} = 0$. We saw something similar to this in Section~2 where we
noticed that the 2-sided ideal generated by the element $\theta$ is the
entire algebra of forms. This second problem is connected with the fact
that we are trying to define the manifold $V$ exactly as a submanifold
in spite of the fact that its `points' are fuzzy and only defined to
within the uncertainty $\mu_P^{-1}$. As a solution to the first problem
we add another dimension and we define $V$ as a submanifold of a
6-dimensional space. We let then $0 \leq i \leq 5$ and we add a
constraint
$$
c_2 = q^5.                                                         \eqno(7.3)
$$
To circumvent the second problem we follow the example furnished by the
quantization of constrained hamiltonian systems. The two constraints are
of second class and Dirac (1964) has shown how in this case one can
introduce a new bracket with respect to which they commute with each
other and with the observables. One can similarly modify the commutation
relations (7.1) so that $c_1$ commutes with the generators of ${\cal A}$
and thus with $c_2$. We introduce the components $r^{ij}$ of an
antisymmetric tensor and we set
$$
[q^i, q^j]^\prime = i \mu_P^{-2} (q^{ij} - r^{ij}).                \eqno(7.4)
$$
From the condition $[q^i, c_2]^\prime = 0$ we find
that 
$$
r^{i5} = q^{i5}.                                                   \eqno(7.5) 
$$
From the condition $[q^i, c_1]^\prime = 0$ we find the equation
$$
q_j r^{ij} + r^{ij} q_j = 2 q^{ij} q_j                             \eqno(7.6)
$$
for $r^{ij}$.  Suppose that the element $q^{i5} q_i$ of ${\cal A}$ is
invertible. The appropriate solution is given then by
$$
r^{ij} = {1 \over 2} q^{[ik} q^{j]5} [q_k, (q^{l5} q_l)^{-1}]_+    \eqno(7.7)
$$
which can also be written in the form
$$
r^{ij} = q^{[ik} q^{j]5} q_k (q^{l5} q_l)^{-1} - {i \over 2} 
\mu_P^{-2} q^{[ik} q^{j]5} q_{kl} q^{l5} (q^{m5} q_m)^{-2}.        \eqno(7.8)
$$

We define ${\cal I}$ to be the 2-sided ideal of ${\cal A}$
generated by the two elements $c_1$ and $c_2$.  By modifying the
original structure of the algebra we have obtained a non-trivial
quotient algebra ${\cal B}$. The modification of the algebra is not
`small' in the sense of the preceding section. We shall consider 
${\cal B}$ as a possible noncommutative equivalent of de~Sitter space.
We refer to Masson (1996) for some examples of quotient algebras
defined without modification of the bracket.

We define $\tilde q_i$ and $\lambda_i$ as in Section 4. The associated
derivations $e_i$ satisfy then
$$
e_i q^j = \delta^j_i - {1 \over 2} [q_i q^{j5} - 
          \delta^5_i q^{jk} q_k, (q^{l5} q_l)^{-1}]_+              \eqno(7.9)
$$
and they no longer commute. They do not even close to form a Lie 
algebra:
$$
[e_i, e_j] = i \mu_P^2 q^{-1}_{ik} q^{-1}_{jl} r^{kl}.            \eqno(7.10)
$$
From the relations
$$
e_i c_1 = 0, \qquad e_i c_2 = 0                                    \eqno(7.11)
$$
it follows that the $e_i$ can be considered as derivations of ${\cal B}$.

If we set
$$
f_i{}^{jk} = \delta^{[j}_{\phantom{[}i} q_{\phantom{[i}}^{k]5} + 
\delta^5_i q^{jk}
$$
then in the commutative limit we can write 
$$
e_i = (q^{l5} q_l)^{-1} f_i{}^{jk} q_k \partial_j.                 \eqno(7.12)
$$
Since $f_i{}^{jk}$ is antisymmetric in the last 2 indices and since 
$f_i{}^{j5} = 0$ the $e_i$ are vector fields on the de~Sitter space.
There are 6 of them and they satisfy 2 relations:
$$
q_i q^{ij} e_j = 0, \qquad q^{i5} e_i = 0.                         \eqno(7.13)
$$

We define a differential calculus as in Section~4.  It follows from
(7.11) and the construction of the differential that
$$
dc_1 = 0, \qquad dc_2 = 0.                                         \eqno(7.14)
$$
The differential calculus can be considered then as a calculus
$\Omega^*({\cal B})$ over ${\cal B}$. It tends as it must in the
commutative limit to the algebra of de~Rham forms over de~Sitter space.
Classical de~Sitter space is parallelizable but it is not obvious that
$\Omega^1({\cal B})$ is free as a left or right module. Because of the
relations (7.13) the $\theta^i$ dual to the $e_i$ cannot be constructed.
Because of the factor $(q^{l5} q_l)^{-1}$ in the expression (7.7) the
equation (2.39) has no solution for the $\lambda_i$ which we have
chosen.

\section{Gauge invariance}

In (5.15) we calculated the connection associated to a perturbation
of the basic differential calculus. One might have expected from the
origin (5.2) of the perturbation that it would resemble rather a
coordinate transformation and that the perturbed connection would
vanish. The fact that this is not the case is due to the existence of
the extra `coordinates' $q^{\mu\nu}$ which have been used to raise and
lower indices.  Our formalism is in fact analogous to that
which can be used to describe a manifold $V$ which is defined by its 
imbedding in a flat space of higher dimension so we mention this
briefly. Let $x^a$ be the coordinates of the imbedding space and
$y^\alpha$ local coordinates of $V$. Then $V$ is defined locally by
functions of the form $x^a(y^\alpha)$.  A variation of $V$ to a surface
$V^\prime$ is given by functions of the form 
$x^{\prime a}(y^\alpha) = x^a(y^\alpha) + h^a(y^\alpha)$.  A variation
of the coordinates of the imbedding space can be written in the form
$x^{\prime a} = x^a + h^a(x^b)$. The most general variation of the
metric on $V$ is obtained by a variation 
$g^\prime_{ab} = g_{ab} + \partial_{(a} h_{b)}$ of the components of the
flat metric. In the commutative limit one can consider 
$(q^\lambda, q^{\mu\nu})$ as the `coordinates' of the imbedding space
and the conditions $q^{\mu\nu} = 0$ the equations which define $V$.

Let $V$ be a smooth manifold and $\phi$ a smooth map of $V$ into itself.
Then $\phi$ induces an automorphism $\phi^*$ of the algebra 
${\cal C}(V)$ of smooth functions given by $\phi^* f = f \circ \phi$ 
and thereby a map $\phi_*$ of the derivations. If $X$ is a derivation of 
${\cal C}(V)$ then so is $\phi_* X = \phi^{*-1} \circ X \circ \phi^*$. 
The noncommutative equivalent of $\phi$ is an automorphism of the algebra
${\cal A}$. Consider the inner automorphism which acts on the generators
by the transformation
$$
q^\lambda \mapsto \ad U^{-1} q^\lambda, \qquad 
q^{\mu\nu} \mapsto \ad U^{-1} q^{\mu\nu}.                           \eqno(8.1)
$$
Then it is obvious that $\lambda_\mu \rightarrow \ad U^{-1} \lambda_\mu$
and a solution to Equation~(3.39) is transformed into another solution
with the same values of $P^{\mu\nu}{}_{\rho\sigma}$, 
$F^\lambda{}_{\mu\nu}$ and $K_{\mu\nu}$. A derivation $X$ is
transformed into $X^\prime = \ad U^{-1} \circ X \circ \ad U$ from which
we deduce that
$$
\theta^\alpha \mapsto \theta^{\prime\alpha} = 
\ad U \circ \theta^\alpha \circ \ad U^{-1}.                         \eqno(8.2)
$$

If the geometry of $V$ is described (locally) by a moving frame
$\theta^\alpha$ then a change of moving frame is a map 
$$
\theta^\alpha \mapsto \theta^{\prime\alpha} = 
\Lambda^\alpha_\beta \theta^\beta                                   \eqno(8.3)
$$
with $\Lambda^\alpha_\beta$ smooth functions of $V$. If the metric is to
be left invariant then there are restrictions on the
$\Lambda^\alpha_\beta$:
$$
\Lambda^\alpha_\gamma \Lambda^\beta_\delta g^{\gamma\delta} =
g^{\alpha\beta}.
$$
In the noncommutative case one could define a change of Stehbein using
the same formula (8.3) but with $\Lambda^\alpha_\beta$
elements of the algebra ${\cal A}$. If 
$\Lambda^\alpha_\beta \in {\cal Z}({\cal A})$ then the new Stehbein has
the same status as the old; it is dual to a set of derivations.
Otherwise the change is purely formal. From the left-linearity of $D^2$
we conclude that
$$
D^2(\Lambda^\alpha_\beta \theta^\beta) = 
\Lambda^\alpha_\beta D^2 \theta^\beta
$$
and so from (3.26) we find that
$$
\Lambda^\alpha_\lambda R^{\prime\lambda}{}_{\beta\gamma\delta} = 
R^\alpha{}_{\lambda\mu\nu} \Lambda^\lambda_\beta 
\Lambda^\mu_\gamma \Lambda^\nu_\delta.                              \eqno(8.4)
$$
If we multiply this equation on both sides by $g^{\beta\gamma}$ and
define the `Ricci tensor' as the map 
$\theta^\alpha \mapsto R^\alpha{}_\delta \theta^\delta$ defined by 
$R^\alpha{}_\delta = g^{\beta\gamma} R^\alpha{}_{\beta\gamma\delta}$
then we find the condition
$$
\Lambda^\alpha_\lambda R^{\prime\lambda}{}_\delta =
R^\alpha{}_\nu \Lambda^\nu_\delta.                                  \eqno(8.5)
$$
The order of the factors is here important. There is no reason for
the trace $R^\alpha{}_\alpha$ of $R^\alpha{}_\delta$ to
be `invariant' except if $\Lambda^\alpha_\beta \in {\cal Z}({\cal A})$.
To define the analogue of the Einstein-Hilbert action one would also
have to introduce a trace on the algebra such that 
$$
\tr (R^\alpha{}_\alpha) \rightarrow \int_V R^\alpha{}_\alpha
$$
in the commutative limit. There is no obvious way in which this can be
done. We are therefore unable at the moment to propose a satisfactory
definition of an action and indeed we are not in a position to argue
that there is even a valid action principle. A discussion of this point
has been made by Connes and coworkers in a series of articles (Kalau \&
Walze 1995, Ackermann \& Tolksdorf 1996, Chamseddine \& Connes 1996) but
the definition which these authors propose is valid only on the
noncommutative generalizations of compact spaces with
euclidean-signature metrics.  Cyclic homology groups have been proposed
(Connes, 1986) as the appropriate generalization to noncommutative
geometry of topological invariants; the appropriate definition of other,
non-topological, invariants in not clear.

\parindent=0cm
\section*{Acknowledgment} One of the authors (JM) would like to thank 
S.  Doplicher and K.  Fredenhagen for interesting conversations. He
would also like to thank A. Dimakis for a critical reading of the
manuscript. This research was partially subsidized by the CEC Science
project No.  SCI-CT91-0729.

\section*{References}

Ackermann T., Tolksdorf J. 1996, {\it A generalized Lichnerowicz
formula, the Wodzicki Residue and Gravity}, Jour. Geom. and Phys. (to
appear).

Bratteli O., Robinson D.W. 1979, {\it Operator Algebras and Quantum 
Statistical Mechanics I}, Springer Verlag.

Chamseddine A., Connes A. 1996, {\it A universal action principle},
IHES Preprint, hep-th/9606056. 

Connes A. 1986, {\it Non-Commutative Differential Geometry}, Publications
of the Inst. des Hautes Etudes Scientifique. {\bf 62} 257.

--- 1994, {\it Noncommutative Geometry}, Academic Press.

Deser S. 1957, {\it General Relativity and the Divergence Problem in
Quantum Field Theory}, Rev. Mod. Phys. {\bf 29} 417.

Dimakis A. 1996, {\it A note on Connections and Bimodules}, q-alg/9603001.

Dimakis A., Madore J. 1996, {\it Differential Calculi and Linear 
Connections}, J. Math. Phys. (to appear).

Dirac  P.A.M. (1964), {\it Lectures on Quantum Mechanics}, 
Yeshiva University Press, New York,

Doplicher S., Fredenhagen K., Roberts, J.E. 1994, 
{\it Spacetime quantization induced by classical gravity}, 
Phys. Lett. {\bf B331} 39.

--- 1995, {\it The Quantum Structure of Spacetime at the Planck Scale 
and Quantum Fields}, Commun. Math. Phys. {\bf 172} 187.

Dubois-Violette M., Michor P. 1996, {\it Connections on Central
Bimodules}, J. Geom. Phys. (to appear).

Dubois-Violette M., Masson T. 1996, {\it On the First-Order Operators
in Bimodules}, Lett. Math. Phys. (to appear).

Dubois-Violette M., Kerner R., Madore J. 1989, {\it Gauge bosons in a
noncommutative geometry}, Phys. Lett. {\bf B217} 485; {\it Classical
bosons in a noncommutative geometry}, Class. Quant. Grav. {\bf 6} 1709.

--- 1990, {\it Noncommutative differential geometry of matrix algebras},
J. Math. Phys. {\bf 31} 316.

--- 1996a, {\it Shadow of Noncommutativity}, Preprint LPTHE Orsay 96/06.

Dubois-Violette M., Madore J., Masson T., Mourad J. 1995, {\it Linear
Connections on the Quantum Plane}, Lett. Math. Phys. {\bf 35} 351.

--- 1996b, {\it On Curvature in Noncommutative Geometry}, 
J. Math. Phys. {\bf} (to appear).

Isham C.J., Salam A., Strathdee J. 1971, {\it Infinity Suppression in 
Gravity-Modified Quantum Electrodynamics}, Phys. Rev. {\bf D3} 1805.

Kalau W., Walze M. 1995, {\it Gravity, Non-Commutative Geometry and the
Wodzicki Residue}, Jour. Geom. and Phys. {\bf 16} 327.

Karoubi M. 1981, {\it Connections, courbures et classes
caract\'eristiques en $K$-th\'eorie alg\'ebrique}, Current trends in
algebraic topology, Part I, London, Ont.

Koszul J.L. 1960, {\it Lectures on Fibre Bundles and Differential Geometry},
Tata Institute of Fundamental Research, Bombay.

Madore J. 1988, {\it Non-Commutative Geometry and the Spinning
Particle}, XI Warsaw Symposium on Elementary Particle Physics,
Kazimierz, Poland; {\it Kaluza-Klein Aspects of Noncommutative
Geometry}, Proceedings of the XVII International Conference on
Differential Geometric Methods in Theoretical Physics, Chester.

--- 1995, {\it An Introduction to Noncommutative Differential Geometry
and its Physical Applications}, Cambridge University Press.

Madore J., Mourad J. 1995, {\it On the origin of Kaluza-Klein structure}
Phys. Lett. {\bf B359} 43.

--- 1996, {\it Noncommutative Kaluza-Klein Theory},
Lecture given at the $5^{\rm th}$ Hellenic School and Workshops on
Elementary Particle Physics, hep-th/9601169.

Masson T. 1996, {\it Submanifolds and quotient manifolds in
noncommutative geometry}, J. Math. Phys. {\bf 37} 2484.

Mourad. J. 1995, {\it Linear Connections in Non-Commutative Geometry},
Class. Quant. Grav. {\bf 12} 965.

Snyder H.S. 1947, {\it Quantized Space-Time}, Phys. Rev. {\bf 71} 38.

\end{document}